\documentclass[fleqn,usenatbib,useAMS]{mnras}
\usepackage{graphicx,slashbox}
\usepackage{amsmath}
\usepackage{amssymb}
\usepackage{bm}
\usepackage{xcolor}
\usepackage{upgreek}

\def\D{\mathrm{d}}
\newcommand{\rhopdf}{$\rho$-PDF}% Density pdf
\newcommand{\Npdf}{$N$-PDF}	% Column density pdf

\title[PDF of a molecular cloud ensemble III]{Density distribution function of a self-gravitating isothermal turbulent fluid in the context of molecular cloud ensembles -- III. Virial analysis}

\author[Donkov et al.]
{\parbox{\textwidth}{S. Donkov$^{1\,\star}$, I. Zh. Stefanov$^2$, T. V. Veltchev$^{3, 4}$ and R. S. Klessen$^{4, 5}$} \vspace{0.4cm} \\
  $^1$Institute of Astronomy and NAO, Bulgarian Academy of Sciences, 72 Tsarigradsko Chausee Blvd., 1784 Sofia, Bulgaria \\
  $^2$Department of Applied Physics, Technical University, 8 Kliment Ohridski Blvd., 1000 Sofia, Bulgaria \\
  $^3$University of Sofia, Faculty of Physics, 5 James Bourchier Blvd., 1164 Sofia, Bulgaria \\
  $^4$Universit\"at Heidelberg, Zentrum f\"ur Astronomie, Institut f\"ur Theoretische Astrophysik, Albert-Ueberle-Str. 2, 69120 Heidelberg, Germany \\
  $^5$Universit\"{a}t Heidelberg, Interdisziplin\"{a}res Zentrum f\"{u}r Wissenschaftliches Rechnen, Im Neuenheimer Feld 205, 69120 Heidelberg, Germany}

\begin{document} 
\label{firstpage}

\date{Accepted 2022 September 14. Received 2022 September 14; in original form 2022 March 8}
\pagerange{\pageref{firstpage}--\pageref{lastpage}} \pubyear{2022}
\maketitle

\begin{abstract}

In the present work we apply virial analysis to the model of self-gravitating turbulent cloud ensembles introduced by Donkov \& Stefanov in two previous papers, clarifying some aspects of turbulence and extending the model to account not only for supersonic flows but for trans- and subsonic ones as well. Make use of the Eulerian virial theorem at an arbitrary scale, far from the cloud core, we derive an equation for the density profile and solve it in approximate way. The result confirms the solution $\varrho(\ell)=\ell^{-2}$ found in the previous papers. This solution corresponds to three possible configurations for the energy balance. For trans- or subsonic flows, we obtain a balance between the gravitational and thermal energy (Case 1) or between the gravitational, turbulent and thermal energies (Case 2) while for supersonic flows, the possible balance is between the gravitational and turbulent energy (Case 3). In Cases 1 and 2 the energy of the fluid element can be negative or zero end thus the solution is dynamically stable and shall be long lived. In Case 3 the energy of the fluid element is positive or zero, i.e., the solution is unstable or at best marginally bound. At scales near the core, one cannot neglect the second derivative of the moment of inertia of the gas, which prevents derivation of an analytic equation for the density profile. However, we obtain that gas near the core is not virialized and its state is marginally bound since the energy of the fluid element vanishes.

\end{abstract} 


\section{Introduction}   \label{Sec-Intr}

Molecular clouds (MCs), the birthplaces of stars, are described by complex physics. Turbulence, gravity, thermodynamics and magnetic fields are regarded to be the main physical agents that determine the structure and the evolution of MCs from their formation in the interstellar medium to their destruction through feedback by the newborn stars \citep{Elme_Scalo_04,HF_12,KG_16}. This complexity can be investigated by various statistical tools which contain various levels of information about the cloud structure and the physics behind it. One appropriate tool is the probability density function (PDF) of column (\Npdf) or mass (\rhopdf) density. The physics and structure of a MC are imprinted, although with some limitations and degeneracies in its PDF. 

Also, the PDF shape may indicate the evolutionary stage of the cloud \citep{Elme_Scalo_04, VS_10, HF_12,KG_16}. At the largest scales and at early evolutionary stages of MCs, supersonic isothermal turbulence dominates their physics which produces a lognormal PDF \citep{VS_94, Federrath_ea_10, Molina_ea_12}. If the cloud mass exceeds its Jeans mass, gravitational collapse starts at larger scales, accompanied by ongoing accretion from the cloud surroundings. This process creates a mass and energy cascade from larger to smaller scales and matter is transferred to small substructures originally formed by supersonic turbulence: clumps and cores. The latter contract in turn as their masses and densities grow \citep[][and the references therein]{VS_ea_19}. At this stage power-law tail(s) (PLTs) emerge and develop at the high-density end of the initially lognormal \rhopdf{} \citep{KNW_11,Girichidis_ea_14,Jaupart_Chabrier_2020}. The PLT regime corresponds to the range of densities of the clumps/cores where star-formation takes place.

The goal of this paper is to investigate the relation between cloud physics and its structure imprinted in the \rhopdf. We focus on advanced stages of MC evolution which correspond to a pronounced PLT of the \rhopdf. In \citet[][hereafter, Paper I]{DS_18} and \citet[][hereafter, Paper II]{DS_19} we elaborated a model based on the notion of {\it average representative object}. It is a model of spherically symmetric and isotropic cloud which is statistically averaged over the so called ``molecular clouds ensemble'' -- the set of all isothermal MCs which have the same \rhopdf, the same temperature, the same size and the same densities at the outer edge and in the centre. The average representative object is the simplest abstract member of the ensemble, with effective size derived from the volume weighted \rhopdf{} in a natural way (see equation \ref{equation_def_scale} in the next Section). This object retains the main structural and physical characteristics of all members of the ensemble (in terms of abstract scales) while the specific morphology and dynamics of each individual are lost. In regard to the dynamics of the average representative object, we assume that the accretion flow is radially symmetric and stationary, i.e., a steady state for the motion of the fluid elements. It has been shown that the steady state is indeed the stable solution for isothermal gas flows accreting radially onto a central core  \citep{Keto_20}.

These simplifications enable us to derive an equation for the \rhopdf{} from the governing equations of the medium: the Euler system (the equation of motion and the continuity equation), the equation of state of the gas and the Poisson equation for the gravitational potential. In addition, the assumption of steady state is consistent with the approximately constant slopes of the PLTs observed at advanced stages of cloud evolution \citep{KNW_11, Girichidis_ea_14, Veltchev_ea_19, Khullar_ea_21}.

The equation of conservation of the total energy, as the sum of the kinetic, thermal and gravitational energy, of the fluid element per unit mass was obtained in Paper I and II from the equations of the medium. Furthermore, through substitution with the explicit form of the energies, one gets a non-linear integral equation for the dimensionless density profile $\varrho(\ell)$ (Paper II). We obtained approximate solutions of this equation in two physical regimes: far and near to the cloud centre. The first solution, characterized by density profile $\varrho(\ell)=\ell^{-2}$, corresponds to balance between the accretion (or, accretion plus turbulent) kinetic energy and the gravitational energy of the cloud's shells located between the fluid element and the central dense region of the cloud, which we labelled ``core''. This corresponds to a steady-state solution, with the fluid element moving inwards with a constant velocity. The second solution corresponds to a free-fall of the fluid element for which the gravitational energy of the core and the accretion kinetic energy balance each other. In this case one gets for the density profile $\varrho(\ell)=\ell^{-3/2}$; this has similarities to the (singular) isothermal sphere solutions \citep[see also, e.g.][]{Whitworth_Summers_85}.

This work is focused on the virial analysis of the suggested model as the latter is extended to account not only for supersonic (turbulent and/or accretion) flows (like in Paper I and II) but for trans- and subsonic ones as well. Applying the Eulerian virial theorem (EVT) \citep{MZ_92, BP_06, Krh_15} to arbitrary radius (scale) in the regime far from the core, under the model assumptions for spherically symmetric cloud, we derive an equation for energy balance per unit mass at this scale. The obtained solution of this equation (equation \ref{equation_EVT-simpler-far-expl}) confirms the profile form $\varrho(\ell)=\ell^{-2}$ in the considered regime and enables us to assess the energy per unit mass of the fluid element, which remained indeterminate in Paper I and Paper II. In regard to the energy balance, we find three possible cases depending on the scaling exponent of the turbulent velocity and on the range of flow velocities in regard to the sonic speed. In two cases the energy of the fluid element can be negative which means that the solution $\varrho(\ell)=\ell^{-2}$ corresponds to a stable state in the outer cloud shells and suggests that these cases must be confirmed from observations. On the other hand, one cannot arrive at an equation for the density profile in the near-to-the-core regime, at least because the second derivative of the moment of inertia of the fluid is not negligible. Qualitative considerations show that the gas around the core is not virialized and it is marginally bound.

The structure of the paper is as follows. Our model is briefly recalled in Section \ref{Sec-Model}. In Section \ref{Sec-Equ_p(s)} we derive the equation for the density profile and obtain an approximate solution. This is the core of our analysis and is presented in four steps: formulation of the Eulerian virial theorem (Section \ref{subsec-EVT}), discussion on its applicability to the suggested model and derivation of the equation for density profile in the far-from-the-core regime in general (Section \ref{subsec-EVT_model}) and explicit form (Section \ref{subsec-EVT_explicit_far}) and obtaining its approximate solution (Section \ref{subsec-EVT_solutions}). Section \ref{Sec-discussion} contains a discussion on our model and the main results. Finally, in Section \ref{Sec-conclusions} we present the conclusions and summarize.

In Table \ref{table_symbols} we provide a list of frequently used symbols and notations in the presented model (Paper I \& II, this work).

\begin{table}
\caption{Frequently used notations in the model}
  \label{table_symbols}
%    \centering
  \begin{tabular}{l@{~~\,}p{6.6cm}}
 \hline 
 \hline
Quantity & Description \\
\hline
$l_{\rm c}$ & Total size of the modeled cloud\\
$\rho_{\rm c}$ & Density at the outer cloud boundary\\
$M_{\rm c}$ & Mass of the cloud\\
$M_{\rm c}^{\ast}$ & Normalizing coefficient ($\lesssim M_{\rm c}$)\\
$l_0$ & Size of the cloud core \\
$\rho_{0},~s_0$  & Density and logdensity of the cloud core\\    
$\ell$ & Dimensionless scale (normalized to $l_{\rm c}$) \\     
$\varrho$ & Dimensionless density (normalized to $\rho_{\rm c}$) \\     
$p$ & Exponent of the power-law density profile \\     
$q$ & Slope of the $\rho$-PDF \\
$c_{\rm s}$ & Sound speed in the cloud\\
$u_{\rm t}$ & Turbulent velocity\\    
$u_{\rm a}$ & Accretion velocity\\    
$u_0$       & Normalizing factor of the turbulent velocity fluctuations in the standard scaling law\\
$\beta$     & Scaling exponent of the turbulent velocity fluctuations \\
$v_{\rm t}$ & Dimensionless turbulent velocity (normalized to $c_{\rm s}$)\\    
$v_{\rm a}$ & Dimensionless accretion velocity (normalized to $c_{\rm s}$)\\    
$\tau$ & Total kinetic plus thermal energy of the fluid\\    
$w$ & Gravitational energy of the cloud \vspace{6pt}\\    
\multicolumn{2}{l}{{\textbf{\textit{Ratios of energies (per unit mass) of the fluid element}}}}\\
$E_0$ & Total energy of a fluid element moving through the cloud scales, vs. thermal energy\\
$G_0$ & Gravitational energy of the entire cloud (excluding the core) vs. thermal energy\\
$G_1$ & Gravitational energy of the core vs. thermal energy, calculated at the outer cloud boundary\\ 
$T_0$ & Turbulent kinetic energy at the outer cloud boundary vs. thermal energy\\
$A_0$ & Accretion kinetic energy at the outer cloud boundary vs. thermal energy\\
$\langle G \rangle$, $\langle T \rangle$, $\langle A \rangle$ & Values of $G_0$, $T_0$ and $A_0$, respectively, averaged over the PDF of the abstract object. \\
\hline
\hline
\end{tabular}
% \smallskip 
\end{table}

\section{On the model}
\label{Sec-Model}

The model presented in Paper I and II is a simplified description of real MCs but, on the other hand, it is also a generalisation, as an average representative of a class of clouds with the same PDFs, sizes, boundary densities and temperatures \citep[see][and Paper I for more details]{DVK_17}. It is not aimed at a detailed description of MC morphology and dynamics but, rather, at reproduction of the main physical properties typical for an entire cloud ensemble. Below we briefly recall it focusing on some aspects of turbulence that need further clarification.

The average representative of the class is modelled by a spherically symmetric, isotropic ball of isothermal \citep{Ferriere_01} and self-gravitating gas with super-, trans- or subsonic turbulent flows therein\footnote{ In Paper I and II we considered only the case of supersonic turbulence and accretion, but here the scope is extended, in view of the obtained possibilities for the energy balance which correspond to the solution for the density profile.}. By use of the PDF one can introduce dimensionless radii of the cloud, which we consider also as spatial scales:
	\begin{equation}
		\label{equation_def_scale}
		\ell(s)=\Bigg(\int\limits_{s}^{s_{0}}p_v(s')\D s'\Bigg)^{1/3}~.
	\end{equation}
Here  $p_v(s)$ is the volume-weighted PDF, $s\equiv\ln(\rho(\ell)/\rho_{\rm c})\equiv\ln(\varrho(\ell))$ is the log-density, where $\rho_{\rm c}$ is the density at the outer boundary of the cloud, $\varrho\equiv\rho/\rho_{\rm c}$ is the dimensionless density. The relation between the mass density profile $\rho(\ell)$ and the PDF\footnote{ Strictly speaking, this relation makes sense only if the density PDF is monotonic -- which is the case considered in the present work.} in differential form is:
	\begin{equation}
		\label{equation_rho(L)-p(s)}
		p_v(s)\D s=-3 \ell^2 \D\ell.
	\end{equation}
The dimensionless radius (scale)  $\ell=l/l_{\rm c}$ varies in the range $\ell_0<\ell\leq 1$, where $l$ is an arbitrary physical scale and $l_{\rm c}$ is the physical scale of the whole cloud. The lower boundary, $\ell_0 \ll 1$, corresponds to the radius of a homogeneous small dense core of density $\rho_0$ and log-density $s_0=\ln(\varrho_0)=\ln(\rho_0/\rho_{\rm c})$ located at the centre of the modelled cloud. 

The model includes also (super-, trans- or subsonic) accretion of matter, which flows from the outer cloud boundary down through all scales to the central core. In addition to the inflow velocity, we also consider turbulent velocity fluctuations which are homogeneous and isotropic at each scale $\ell$ in the interval $\ell_0<\ell\leq 1$ presumed to lie within the inertial range of the turbulent cascade where dissipation can be neglected. Therefore it is convenient to use velocity decomposition $\vec{u}=\vec{u}_{\rm a}+\vec{u}_{\rm t}$ with $\vec{u}_{\rm a}=-u_{\rm a}\hat{e}_r$ (only infall) and $\vec{u}_{\rm t}$ being isotropic. The entire cloud is at steady state at each scale both in regard to the macro-states (the motion of the fluid elements) and the micro-states (the thermal motion of the molecules). We consider it as averaged over an ensemble of spherical objects (clouds) which have the same PDFs (also sizes, boundary densities and temperatures) and dynamics as described above. Due to the assumption of steady state one can apply the ergodic hypothesis and hence substitute the averaging with regard to time for averaging over the ensemble of copies of our system which describes all dynamical states of the fluid elements. Let us then consider, at a fixed moment in time, a fluid element which resides in a given fluid volume of the cloud, at scale $\ell$. The averaged velocity field is equal to the infall velocity $\vec{u}_{\rm a}=-u_{\rm a}\hat{e}_r$, because the averaged turbulent velocity vanishes due to the assumption of isotropy of the turbulent motions\footnote{ After the averaging, the turbulent velocity does not contribute to the motion of the fluid element. However, its contribution to the kinetic energy term $\langle u^2 \rangle$ is non-vanishing due to the scalar nature of the latter. The velocity is first squared, then averaged.}. The density field must be averaged too which yields a corresponding density profile -- if the PDF is of power-law type (i.e. monotonically decreasing function of density).

The model was further developed in Paper II to account for the gravity of matter external in regard to the cloud. In the spirit of our model, we assume that the material outside the cloud boundary obeys a radial symmetry and gives rise to a potential $\varphi^{\rm ext}$ in the cloud volume, where it is constant and does not contribute to the gravitational force acting on the fluid elements.
		
\section{Derivation of the equations for the mass density profile from Eulerian Virial Theorem}   
\label{Sec-Equ_p(s)}

\subsection{Eulerian Virial Theorem}
\label{subsec-EVT}

Let us briefly consider the Eulerian virial theorem (EVT), that is the virial theorem for a fixed volume $V$, surrounded by a surface $S$. We focus on the simplified scalar form of the tensorial virial theorem which follows from the Boltzmann equation. For the derivation of this form and comments on the physical nature of its terms we refer the reader to \cite{MZ_92}, \cite{BP_06} and \cite{Krh_15}.

The use of EVT, instead of Lagrangian virial theorem (LVT), is essential for the treatment in this work because only the EVT allows for the description of a non-equilibrium but stationary system. Since we consider a steady-state gas flow through mass shells at a given radius (and not the inward motion of the shells themselves) we naturally operate from Eulerian perspective. The EVT refers to fixed volume elements in space, through which the gas can flow, rather than to fixed mass elements which necessarily change their location as the gas flows. Moreover, in case the second time derivative of the moment of inertia is zero, the system in the LVT treatment is necessarily in equilibrium, or in a ballistic trajectory at most, with zero net forces (see the discussion in \citealt{BP_06}), whereas in the EVT treatment it is either in static equilibrium without any inside flows (e.g., gravity is balanced by gas pressure), or in steady state. Thus, the application of the EVT to a cloud in steady state allows for obtaining profiles of density and infall speed which depend only on radius, but not on time.

 The mathematical expression of the EVT is:

\begin{equation}
\label{equation_EVT}
\frac{1}{2} \ddot{I} = 2\big( \tau-\tau_{S} \big) + w - \frac{1}{2}\frac{\D}{\D t}\int\limits_{S}\rho r^2 \vec{u} \cdot \vec{\D S}~,
\end{equation}
where $\vec{r}$ is a radius-vector (with absolute value $r=\lvert\vec{r}\rvert$) from the origin of the Eulerian coordinate system, $\vec{u}(\vec{r},t)$ is the velocity field (with absolute value $u=\lvert\vec{u}\rvert$) in the fluid and $\rho(\vec{r},t)$ is the density field. The terms in equation (\ref{equation_EVT}) have the explicit form and physical meaning as follows:

\begin{equation}
\label{equation_I}
I= \int\limits_{V} \rho r^2 \D V~,
\end{equation}
is the moment of inertia of the fluid in the volume $V$, calculated about the origin, and $\ddot{I}$ is its second time derivative;

\begin{equation}
\label{equation_tau}
\tau= \int\limits_{V} \bigg( \frac{1}{2} \rho u^2 + \frac{3}{2} P_{\rm th} \bigg) \D V~,
\end{equation}
is the total kinetic plus thermal energy of the fluid ($P_{\rm th}$ is the thermal pressure);

\begin{equation}
\label{equation_T_S}
\tau_{S}= \frac{1}{2} \int\limits_{S} \vec{r}\cdot \Pi \cdot\vec{\D S}~,
\end{equation}
is the confining pressure on the volume surface, including both the thermal pressure and the ram pressure of any gas flowing across the surface; where $\Pi_{ij}=\rho u_{i}u_{j}+P_{\rm th}\delta_{ij}$ is the general form of the fluid pressure tensor;
\begin{equation}
\label{equation_W}
w= - \int\limits_{V} \rho \vec{r} \cdot \vec{\nabla} \phi \,\D V~,
\end{equation}
is the gravitational energy of the cloud, where the potential $\phi$ accounts both for self-gravity of the cloud and gravity of the external matter. The last term on the r.h.s. in equation (\ref{equation_EVT}) represents the rate of change of the momentum flux across the cloud surface.

We note that we neglect the magnetic fields in our model (see Paper I, Section 3.1), in order to simplify the first step to obtain the density profile of the fluid from first principles.

\subsection{Applicability of the EVT to our model: general consideration}
\label{subsec-EVT_model}

The EVT is a powerful tool as it describes the global energy balance in the objects/systems to which it is applied. We use the EVT for obtaining information about the energy balance for the entire cloud as well as for an arbitrary scale $\ell$ in its interior. Moreover, the modelled cloud is spherically symmetric and, with the origin of the Eulerian coordinate system set in the centre, the terms of the EVT are easy to calculate -- in view of the radial accretion flow and the assumed steady state for macro- and micro-states. We stress that all terms in equation (\ref{equation_EVT}) are assumed to be ensemble averaged as described in Paper I (Section 3.1) and in Section \ref{Sec-Model} above\footnote{All terms are averaged in respect to the micro-states determined by the turbulence at a given scale $\ell$. The averaged terms are put in $\langle...\rangle$.}. 

There are two points to comment on the application of the EVT to the cloud considered here, i.e. to scales $\ell_0<\ell\leqslant1$.

First, one can see that the last term in equation (\ref{equation_EVT}) becomes zero (for $\ell=1$ as well) since $\langle \rho r^2 \vec{u} \rangle = {\rm const}(t)$ due to the steady-state condition.

Second, let us consider the term $\langle \ddot{I} \rangle$. The model assumes that -- despite the accretion flow through the scales, -- no material is being accumulated in the cloud shells; all the gas eventually flows into the cloud core. The latter is presupposed to be very small: of nearly constant size and density within the characteristic accretion time. Therefore, considering a scale $\ell$ far from the core ($\ell\gg\ell_0$), one can suppose $\langle \ddot{I} \rangle \simeq 0$ and equation (\ref{equation_EVT}) takes the form:

\begin{equation}
	\label{equation_EVT-simpler-1}
	0 = 2\big(\langle \tau \rangle - \langle \tau_{S} \rangle \big) + \langle w \rangle ~.
\end{equation}

On the other hand, in the regime near to the core ($\ell \gtrsim \ell_0$) one should be cautious and not adopt $\langle \ddot{I} \rangle \simeq 0$, regardless of the presuppositions in the model.

\subsection{The explicit form of the EVT equation}
\label{subsec-EVT_explicit_far}

In this Section we derive the explicit form of equation (\ref{equation_EVT-simpler-1}) in the framework of the model (Section \ref{Sec-Model}) and using the explicit form of the terms $\langle \tau \rangle$ (formula \ref{equation_tau}), $\langle \tau_S \rangle$ (formula \ref{equation_T_S}), and $\langle w \rangle$ (formula \ref{equation_W}). 

\subsubsection{The explicit form of $\langle \tau \rangle$}
\label{subsubsec-<T>_explicit}

It is convenient to normalize all terms in equation (\ref{equation_EVT-simpler-1}) to the thermal energy per unit mass: $c_{\rm s}^2$. Introducing a dimensionless velocity field $v\equiv u/c_{\rm s}$, the kinetic energy term can be written (see Paper I, Sections 3.2 and 3.3) as:
	
\begin{equation}
	\label{equation_rhov=rhovt+rhova}
	\langle \varrho v^2 \rangle =\varrho\langle v^2 \rangle=\varrho\langle v_{\rm t}^2 \rangle + \varrho\langle v_{\rm a}^2 \rangle~,
\end{equation}
where $\varrho\langle v_{\rm t}^2 \rangle$ is the kinetic energy per unit volume due to turbulent velocity fluctuations and $\varrho\langle v_{\rm a}^2 \rangle$ is the kinetic energy per unit volume due to accretion of matter towards the cloud centre. 
	
Since the considered spherically symmetric cloud is ensemble averaged we choose to apply a standard scaling relation for  $\langle v_{\rm t}^2 \rangle$ (see Paper II, Section 2.1):
	\begin{eqnarray}
	\label{equation_vt2}
	\langle v_{\rm t}^2 \rangle = \frac{u_0^2}{c_{\rm s}^2}\bigg(\frac{l_{\rm c}}{1~{\rm pc}}\bigg)^{2\beta} \ell^{2\beta} = T_0 \ell^{2\beta}~,
	\end{eqnarray}
where $u_0$ and $0\lesssim\beta\lesssim1$ are the normalizing factor and the scaling exponent of the turbulent velocity fluctuations in the standard scaling law:

\begin{equation}
	\label{equation_vt}
	u=u_0 L^{\beta} ~,
\end{equation} 
 
\citep{K41, Larson_81, Pad_ea_06, Kritsuk_ea_07, Federrath_ea_10}, and $T_0\equiv (u_0^2/c_{\rm s}^2)(l_{\rm c}/{1~{\rm pc}})^{2\beta}$ is the ratio of the turbulent kinetic energy per unit mass of the fluid element at the cloud boundary to the thermal energy per unit mass. The explicit form of $\langle v_{\rm t}^2 \rangle$ presupposes that the turbulence is formally independent of the accretion. 
	
The explicit form of the accretion term can be obtained from the continuity equation:
\begin{equation}
	\label{equation_va2}
	\langle v_{\rm a}^2\rangle = A_0 \varrho(\ell)^{-2} \ell^{-4}~,
% 	=\ell^{4}\varrho(\ell)^{2} \langle v_{\rm a}^2\rangle
	\end{equation}
where $A_0={\rm const}(\ell)$ (see Paper I, Section 3.3). Since the quantities are dimensionless and normalized, one can obtain $A_0$ choosing $\ell=1$, i.e. at the cloud boundary. Then $\varrho=1$ and one gets $\langle v_{\rm a}^2\rangle = u_{\rm a,c}^{2}/c_{\rm s}^{2}$: the ratio of the accretion kinetic energy term at the cloud boundary to the thermal kinetic energy per unit mass.
	
Now, taking into account the presuppositions in Section \ref{Sec-Model} and assuming a density profile for the cloud in the form $\varrho(\ell)=\ell^{-p}$ and the normalized thermal energy per unit volume $P_{\rm th}= \varrho(\ell)=\ell^{-p}$, the term $\langle \tau \rangle$ can be calculated as follows:
	
	\begin{eqnarray}
	\label{equation_<T>}
	\langle \tau \rangle = 3M_{\rm c}^{*} c_{\rm s}^2 \int\limits_V \left\langle \frac{1}{2}\varrho v^2 + \frac{3}{2} P_{\rm th} \right\rangle \frac{\D V}{4\uppi l_{\rm c}^3} = \nonumber \\
	3M_{\rm c}^{*} c_{\rm s}^2 \int\limits_{\ell_0}^{\ell} \bigg( \frac{1}{2} T_0 \ell'^{(-p+2\beta)} + \frac{1}{2} A_0 \ell'^{(p-4)} + \frac{3}{2}\ell'^{(-p)} \bigg) \ell'^{2} \D\ell' \nonumber \\
	 + \langle \tau \rangle_{0} = \nonumber \\
% 	= ... \nonumber \\
	= 3M_{\rm c}^{*} c_{\rm s}^2 \bigg[\frac{1}{2}T_0 \frac{\ell^{2\beta+3-p} - \ell_0^{2\beta+3-p}}{2\beta+3-p} + \frac{1}{2}A_0 \frac{\ell^{p-1} - \ell_0^{p-1}}{p-1} + \nonumber \\
	 + \frac{3}{2}\frac{\ell^{3-p} - \ell_0^{3-p}}{3-p}\bigg] + \langle \tau \rangle_{0}~. 
	\end{eqnarray}
The normalizing coefficient $M_{\rm c}^{*}=(4/3)\uppi \rho_{\rm c} l_{\rm c}^3$ is lower than the cloud mass by a factor of few; it is the mass of a ball with radius equal to the radius of the cloud boundary and with homogeneous density equal to the density at this boundary. The term $\langle \tau \rangle_{0}$ denotes the kinetic and thermal energy of the core. Its exact expression requires additional assumptions about the physics of the core -- at least, its thermodynamic state and the velocity field in it. We avoid such complication of the model and note only that one should expect that $\langle \tau \rangle_{0} \sim \ell_0^{\alpha}$, where the exponent $\alpha$ is positive and of order unity\footnote{We remind the reader that the core is homogeneous with density $\rho_0\propto \ell_0^{-p}$ (Section \ref{Sec-Model}). Then, on the general assumption of polytropic equation of state, the thermal energy of the gas is $\sim \rho_0^\Gamma l_0^3\propto\ell_0^{3-p\Gamma}$, with $\Gamma\sim 1$. In addition, if turbulence is the sole contributor to the kinetic energy the latter is $\sim \rho_0 l_0^{3+2\gamma}\propto\ell_0^{3+2\gamma-p}$, where $\gamma\leqslant 1$ is the turbulent velocity scaling exponent in the core.}. Considering the terms in equation (\ref{equation_<T>}) which contain explicitly $\ell_0$ to some power $\delta$, one can notice that the exponent $\delta\sim 1$ since observational and numerical studies show that $2 \gtrsim p \gtrsim 1$ \citep{Veltchev_ea_19,VS_ea_19} and $1 \gtrsim \beta \gtrsim 0$ \citep{Goodman_1998,Elme_Scalo_04,Shetty_2012,KG_16}. 

In view of the relations $\ell\gg\ell_0$ and $\alpha\sim\delta\sim 1$, all terms in equation (\ref{equation_<T>}) containing $\ell_0^{\delta}$ ($0<\delta\sim 1$) and $\ell_0^{\alpha}$ ($0<\alpha\sim 1$) can be neglected in comparison to the ones containing $\ell^{\delta}$ ($0<\delta\sim 1$). From these considerations we arrive at a more simple form for $\langle \tau \rangle$:
	
\begin{equation}
	\label{equation_<T>_proxy}
	\langle \tau \rangle \simeq 3M_{\rm c}^{*} c_{\rm s}^2 \bigg[ \frac{1}{2}T_0 \frac{\ell^{2\beta+3-p}}{2\beta+3-p} + \frac{1}{2}A_0 \frac{\ell^{p-1}}{p-1} + \frac{3}{2}\frac{\ell^{3-p}}{3-p} \bigg] ~.
\end{equation}

\subsubsection{The explicit form of $\langle \tau_S \rangle$}
\label{subsubsec-<T_S>_explicit}
	
Let us first consider the integrand in equation (\ref{equation_T_S}) in dimensionless form, according to the model assumptions:

\begin{eqnarray}
	\label{integrand tau_S}
	\langle \vec{\ell}\cdot \Pi \cdot\vec{\D S} \rangle = \nonumber\\
\big[  \varrho \langle v_{\rm a}^2 \rangle + \varrho \langle v_{\rm t}^2 \rangle \langle \cos\alpha^2 \rangle + 2 \varrho \langle v_{\rm a} v_{\rm t} \rangle \langle \cos\alpha  \rangle +  \varrho \big] \nonumber\\
\ell^3 \sin\theta \D \theta \D \varphi ~,
\end{eqnarray}

\noindent{}where $\vec{\D S} = \hat{\ell} \ell^2 \sin\theta \D \theta \D \varphi$, $\hat{\ell}$ is unity vector pointing outwards, $\theta$ is the polar angle in the range $[0,\pi]$ and $\varphi$ is the azimutal angle in the range $[0,2\pi]$. The last addend in equation (\ref{integrand tau_S}) $\varrho$ originates from the thermal pressure term. According to the model, the tensor product in $\Pi$ can be presented as follows: $\vec{v}\otimes\vec{v} = \vec{v_{\rm a}}\otimes\vec{v_{\rm a}} + \vec{v_{\rm t}}\otimes\vec{v_{\rm t}} + \vec{v_{\rm a}}\otimes\vec{v_{\rm t}} + \vec{v_{\rm t}}\otimes\vec{v_{\rm a}}$. Hence

$$\vec{\ell}\cdot\vec{v}\otimes\vec{v}\cdot\hat{\ell} = \ell v_{\rm a}^2 + \ell v_{\rm t}^2 \cos\alpha^2 + 2 \ell v_{\rm a} v_{\rm t} \cos\alpha ~,$$

\noindent{}where $\vec{\ell}\cdot\vec{v_{\rm t}} = \ell v_{\rm t} \cos\alpha$, $\hat{\ell}\cdot\vec{v_{\rm t}} = v_{\rm t} \cos\alpha$, and $\alpha$ is an accidental angle which takes values in the range $[0,\pi]$ with equal probability, due to the assumption of isotropic turbulence. Also, one obtains for the accretion velocity $\vec{\ell}\cdot\vec{v_{\rm a}} = - \ell v_{\rm a}$ and $\hat{\ell}\cdot\vec{v_{\rm a}} = - v_{\rm a}$ because of the opposite directions of the vectors $\vec{\ell}$ and $\hat{\ell}$, on the one hand, and of $\vec{v_{\rm a}}$, on the other hand.

In view of the statistical nature of $\alpha$ one sees easily that:

$$ \langle \cos\alpha  \rangle = \frac{1}{\pi} \int_{0}^{\pi} \cos\alpha \D \alpha = 0 ~, $$

$$ \langle \cos\alpha^2  \rangle = \frac{1}{\pi} \int_{0}^{\pi} \cos\alpha^2 \D \alpha = \frac{1}{2} ~. $$

The integration over the sphere with radius $\ell$ in equation (\ref{equation_T_S}) yields simply a factor $4\pi$, because there is no any addend in the parentheses in equation (\ref{integrand tau_S}) which depends on $\theta$ or $\varphi$. Thus one arrives at:

$$\langle \tau_S \rangle \sim 4\pi \bigg( \ell^3 \varrho \langle v_{\rm a}^2 \rangle + \frac{1}{2} \ell^3 \varrho \langle v_{\rm t}^2 \rangle + \ell^3\varrho \bigg) ~.$$

Finally, plugging in the density profile of the presumed type $\varrho (\ell)=\ell^{-p}$ and the expressions for $\langle v_{\rm t}^2 \rangle$ and $\langle v_{\rm a}^2 \rangle$ from equations (\ref{equation_vt2}) and (\ref{equation_va2}), and normalizing in respect to density, velocity and scale (like in Section \ref{subsubsec-<T>_explicit}), we obtain the explicit form of the surface term in the EVT as follows:

\begin{equation}
	\label{equation <tau_S> explicit}
	\langle \tau_S \rangle = 3M_{\rm c}^{*} c_{\rm s}^2 \bigg[ \frac{1}{2} A_0 \ell^{p-1} + \frac{1}{4} T_0 \ell^{2\beta+3-p} + \frac{1}{2} \ell^{3-p} \bigg] ~.
\end{equation}

\subsubsection{The explicit form of $\langle w \rangle$}
\label{subsubsec-<W>_explicit}
	
The gravitational term $\langle w \rangle$ is calculated in the EVT equation as follows. Note that $-\vec{\nabla} \phi$ in equation (\ref{equation_W}) is the gravitational force per unit mass acting on a fluid element. Due to the Gau\ss{} theorem, in spherical systems only mass inside a sphere with radius $\ell$ contributes to the force, and we can write:
	
	\begin{equation}
	\label{equation_grav-force}
	-\vec{\nabla}\phi= c_{\rm s}^2 \bigg[-\frac{G}{c_{\rm s}^2}\frac{M(\ell)}{l_{\rm c}^2}\frac{\vec{\ell}}{\ell^3} - \frac{G}{c_{\rm s}^2}\frac{M_0}{l_{\rm c}^2}\frac{\vec{\ell}}{\ell^3}\bigg]~,
	\end{equation}
where $\vec{\ell}$ is a dimensionless vector with absolute value $\ell$ and directed radially outwards, $G$ is the gravitational constant, $M_0$ is the core mass, $$M(\ell)=3M_{\rm c}^{*}\int\limits_{\ell_0}^{\ell} \ell'^2\varrho(\ell')d\ell' = \frac{3}{3-p}M_{\rm c}^{*}(\ell^{3-p}-\ell_0^{3-p}) $$ is the mass of the inner shells in respect to $\ell$, and $M_{\rm c}^{*}= (4/3)\uppi l_{\rm c}^3 \rho_{\rm c}$ is the normalizing coefficient (see Section \ref{subsubsec-<T>_explicit}). Formula (\ref{equation_grav-force}) can be rewritten in a more simple form recalling the dimensionless coefficients $G_0=(2G/c_{\rm s}^2)(M_{\rm c}^{*}/l_{\rm c})$ and $G_1=(2G/c_{\rm s}^2)(M_0/l_{\rm c})$ introduced in Paper I (Section 3.4) and evaluated for the outer cloud boundary $l_{\rm c}$. They are interpreted, respectively, as the ratio of the gravitational energy of the entire cloud (excluding the core) to its thermal energy per unit mass ($G_0$) and the ratio of the gravitational energy per unit mass of the core to the thermal energy per unit mass ($G_1$). 

Now the equation (\ref{equation_grav-force}) is transformed to:

	\begin{equation}
\label{equation_grav-force-G}
-\vec{\nabla}\phi= (c_{\rm s}^2/l_{\rm c}) \bigg[-\frac{1}{2} G_0 \frac{3}{3-p} (\ell^{3-p}-\ell_0^{3-p}) \frac{\vec{\ell}}{\ell^3} - \frac{1}{2} G_1 \frac{\vec{\ell}}{\ell^3}\bigg]~,
\end{equation}
and, after the corresponding substitutions in equation (\ref{equation_W}), one gets:

\begin{eqnarray}
\label{equation_<W>1}
\langle w \rangle = - 3M_{\rm c}^{*} \int\limits_{V} \frac{\rho}{\rho_{\rm c}} \vec{l} \cdot \vec{\nabla}\phi \frac{\D V}{4\uppi l_{\rm c}^3} = \nonumber \\
= 3M_{\rm c}^{*} c_{\rm s}^2 \int\limits_{\ell_0}^{\ell} \ell'^{(-p)} \vec{\ell'} \cdot \bigg( -\frac{1}{2} G_0 \frac{3}{3-p} (\ell'^{(3-p)}-\ell_0^{3-p}) \frac{\vec{\ell'}}{\ell'^3} - \nonumber \\
- \frac{1}{2} G_1 \frac{\vec{\ell'}}{\ell'^3} \bigg) \ell'^2 \D\ell' + \nonumber \\
+ 3M_{\rm c}^{*} c_{\rm s}^2 \int\limits_{0}^{\ell_0} \ell_0^{(-p)} \vec{\ell'} \cdot \bigg( -\frac{1}{2} G_0 \frac{3}{3-p} (\ell'^{(3-p)}-\ell_0^{3-p}) \frac{\vec{\ell'}}{\ell'^3} - \nonumber \\
- \frac{1}{2} G_1 \frac{\vec{\ell'}}{\ell'^3} \bigg) \ell'^2 \D\ell'~.
\end{eqnarray}
 Note that in the second integral we take into account the assumption that the core is homogeneous and thus the density profile is simply $\varrho=\ell_0^{-p}={\rm const}(\ell)$. After some algebra we arrive at:
 
 \begin{eqnarray}
 \label{equation_<W>2}
 \langle w \rangle = 3M_{\rm c}^{*} c_{\rm s}^2 \bigg[ - \frac{3}{3-p} \frac{G_0}{2} \bigg( \frac{\ell^{5-2p}-\ell_0^{5-2p}}{5-2p} - \nonumber \\ \frac{\ell^{2-p}-\ell_0^{2-p}}{2-p} \ell_0^{3-p}  + \frac{p-3}{2(5-p)}\ell_0^{5-2p} \bigg) - \nonumber \\
 - \frac{G_1}{2} \bigg( \frac{\ell^{2-p}-\ell_0^{2-p}}{2-p} + \frac{\ell_0^{2-p}}{2} \bigg) \bigg]~.
 \end{eqnarray}
 
 The same reasoning about sub-terms of the kinetic energy term $\langle \tau \rangle$ containing $\ell^{\delta}$ or $\ell_0^{\delta}$ ($0<\delta\sim 1$) presented in Section \ref{subsubsec-<T>_explicit} is applicable for the gravitational energy term $\langle w \rangle$ as well. One can neglect the terms containing $\ell_0^{\delta}$ and retain the leading ones with $\ell^{\delta}$ because $\ell\gg\ell_0$. Then equation (\ref{equation_<W>2}) is simplified to the form:
 
 \begin{eqnarray}
 \label{equation_<W>2_proxy}
 \langle w \rangle \simeq 3M_{\rm c}^{*} c_{\rm s}^2 \bigg[- \frac{3}{3-p} \frac{G_0}{2} \frac{\ell^{5-2p}}{5-2p} - \frac{G_1}{2} \frac{\ell^{2-p}}{2-p} \bigg] ~.
 \end{eqnarray}
 
Eventually, one obtains equation (\ref{equation_EVT-simpler-1}) in an explicit normalized form:
 
 \begin{eqnarray}
 \label{equation_EVT-simpler-far-expl}
 \frac{[2\langle \tau \rangle - 2\langle \tau_S \rangle + \langle w \rangle]}{3M_{\rm c}^{*} c_{\rm s}^2} \simeq \nonumber \\
 \bigg( \frac{1}{2\beta+3-p} - \frac{1}{2} \bigg) T_0 \ell^{2\beta+3-p} + \bigg( \frac{1}{p-1} - 1 \bigg)A_0 \ell^{p-1} +  \nonumber \\
 \bigg( \frac{3}{3-p} - 1 \bigg) \ell^{3-p} - \frac{3}{3-p} \frac{G_0}{2} \frac{\ell^{5-2p}}{5-2p} - \frac{G_1}{2} \frac{\ell^{2-p}}{2-p} \simeq 0 ~.
 \end{eqnarray}

\subsection{Solutions of equation (\ref{equation_EVT-simpler-far-expl})}
\label{subsec-EVT_solutions}

To obtain possible solutions of equation (\ref{equation_EVT-simpler-far-expl}) we use the approach of comparing the exponents of its terms. The method was already used in Papers I and II and explained in details in Section 3.1 of Paper II. Here we briefly recall the main idea, for clarity.

After the approximations discussed in the previous Section, the EVT equation (\ref{equation_EVT-simpler-far-expl}) contains only terms with power-law dependence on $\ell$. Since $0< \ell \leq 1$ the terms with lower powers dominate over the higher ones. The idea of the approach is to retain only the dominant terms (of the lowest power) and to neglect the remaining ones. {The dominant terms must be at least two; a single dominant term would remain unbalanced and then the only solution would be the trivial one. The powers of $\ell$ factor out and only constants remain. At the end, one obtains an equation which gives the balance between the different energy terms.

To identify the leading order terms and to obtain an equation for $p$, one chooses a pair of terms with the hypothesis that they are dominant and have equal powers. The equality of the powers results in a simple equation for $p$. Adopting the obtained root for $p$, the values of all exponents in the equation are evaluated and the validity of the considered hypothesis is confirmed or rejected. The same recipe is applied to all possible pairs of terms \citep{Zhivkov_99, RHB_2006}.   

In our model the exponents of the terms in equation (\ref{equation_EVT-simpler-far-expl}) are: $p-1,~3-p,~2\beta+3-p,~5-2p$ and $2-p$. The roots for $p$ obtained by the algorithm described above are: $2,~3/2,~3,~2+\beta,2-2\beta$, respectively (see Table \ref{tab-1}) and the values of the exponents for each of them are given in Table \ref{tab-2}. We remind the reader that $0\lesssim\beta\lesssim1$ (Section \ref{subsubsec-<T>_explicit}). 

As discussed in Section \ref{subsec-EVT_model}, there are two different physical regimes in the framework of our cloud model: far from the core and near to the core. Below we explore the solutions for the density profile in the former case and comment on the difficulty to obtain such in the regime near to the core.

\subsubsection{The regime far from the core}
\label{subsubsec-EVT_solution_far}

Here one neglects the last term in equation (\ref{equation_EVT-simpler-far-expl}), i.e. the core contribution in the gravitational term. This reduces the set of exponents to $p-1,~3-p,~2\beta+3-p$ and $5-2p$ and yields roots for $p$: 2, $2+\beta$ and $2-2\beta$. Now there are two possibilities for the energy balance, depending on the turbulent velocity scaling:

(i) If $\beta>0$, there is a single solution for $p=2$ and the energy balance is in the form: 
\begin{equation}
\label{equation_solution_far_from_the_core_A}
  4 - 3G_0 \approx 0 ~.
\end{equation}

(ii) If $\beta=0$, there are three equal roots, corresponding to the single solution $p=2$, whereas the energy balance is different: 
\begin{equation}
\label{equation_solution_far_from_the_core_B}
  T_0 + 4 - 3G_0 \approx 0 ~.
\end{equation}

In case of a positive turbulent velocity scaling exponent (i) gravity is balanced by thermal energy; the accretion and turbulent terms do not play any role. Apparently such a balance is possible only if the accretion and turbulent velocities are of the order of the sound speed or less. This subcase yields $G_0\approx 4/3$. The result for the energy balance of the fluid element per unit mass from Paper II is not applicable here, since it was obtained only for supersonic (accretion and/or turbulent) flows. Hence one is not able to obtain the energy $E_0$ of fluid element.

On the contrary, if the turbulent velocity does not depend on the spatial scale (ii), there exist two further possibilities. If the accretion and turbulent velocities are of the order of the sound speed or less, the energy balance reads: $T_0 + 4 - 3G_0 \approx 0$; then the results from Paper II are inapplicable again (the energy $E_0$ of the fluid element remains undetermined). But if the turbulent flow is supersonic, the balance is: $T_0 - 3G_0 \approx 0$, and the result from Paper II is valid: $A_0 + T_0 - 3G_0 \approx E_0$. Combining these two equations, one obtains that $E_0 \approx A_0 > 0$. The instance $E_0 \approx 0$ is also possible -- if the turbulent flow is supersonic while the accretion one is trans- or subsonic; then the accretion term does not play a role in the equation (see Paper II, Sec. 4, forth paragraph). Hence, the energy of fluid element is positive (an unstable energy state), if both the turbulence and accretion are supersonic, and is roughly zero (a marginally bound energy state), if only the turbulence is supersonic.

\begin{table}
 	\centering
 	\small
 	\caption{Possible values of the density profile exponents $p$ obtained from comparison between the exponents of the terms in the equation (\ref{equation_EVT-simpler-far-expl}).}
 	\begin{tabular}{|c|c|c|c|c|c|}  
		\hline
		\hline
		exponents     & $p-1$     & $3-p$    & $2\beta+3-p$   & $5-2p$       & $2-p$ \\
		\hline
		$p-1$         & --        & $2$      & $2+\beta$      & $2$          & $3/2$ \\
		\hline
		$3-p$         & $2$       & --       & --             & $2$          & --    \\
		\hline
		$2\beta+3-p$  & $2+\beta$ & --       & --             & $2-2\beta$   & --    \\
		\hline
		$5-2p$        & $2$       & $2$      & $2-2\beta$     & --           & $3$   \\
		\hline
		$2-p$         & $3/2$     & --       & --             & $3$          & --    \\
		\hline
		\hline
	\end{tabular}\label{tab-1}
\end{table}

\begin{table}
	\centering
	\small
	\caption{Calculated exponents of the terms in equation (\ref{equation_EVT-simpler-far-expl}), corresponding to the roots for the density profile $p$ from Table \ref{tab-1}.}
	\begin{tabular}{|l@{\,}|c@{\,}|c@{\,}|c|c|c|}  
		\hline
		\hline
		\backslashbox{$p$}{exp.} 	 & $p-1$        & $3-p$        & $2\beta+3-p$     & $5-2p$       & $2-p$     \\
		\hline
		$2$                                  & $1$          & $1$          & $2\beta+1$       & $1$          & $0$      \vspace{4pt} \\
		$3/2$                                & $1/2$        & $3/2$        & $2\beta+3/2$     & $2$          & $1/2$    \vspace{4pt} \\
		$3$                                  & $2$          & $0$          & $2\beta$         & $-1$         & $-1$      \vspace{4pt} \\
		$2+\beta$                            & $\beta+1$    & $1-\beta$    & $\beta+1$        & $1-2\beta$   & $-\beta$  \vspace{4pt} \\
		$2-2\beta$                           & $1-2\beta$   & $1+2\beta$   & $1+4\beta$       & $1+4\beta$   & $2\beta$  \\
		\hline
		\hline
	\end{tabular}\label{tab-2}
\end{table}

Our model allows to write down the energy-balance equations as the different forms of energy are averaged over the PDF of the abstract object. Let us recall the definition of $G_0$ and of the average density of the cloud which in the case of a long PL-tail can be evaluated from $\langle\rho\rangle_{\rm c}= \rho_{\rm c} q/(1+q)$ $~$ \citep{DVK_17}, where $q$ is the PDF slope and $q=-3/p$. Then in the case $q=-3/2$ ($p=2$) the above formula gives $\langle\rho\rangle_{\rm c}= 3\rho_{\rm c}$, and, since $G_0=(2G/c_{\rm s}^2)(M_{\rm c}^{*}/l_{\rm c})\sim \rho_{\rm c}$, we can obtain the average, over the entire cloud (our abstract object), gravitational energy per unit mass of the fluid element: $\langle G \rangle\sim \langle\rho\rangle_{\rm c}$, hence $\langle G \rangle= 3G_0$. For $p=2$ the accretion energy per unit mass remains constant from one scale to the other: $\langle v_{\rm a}^2\rangle\propto \varrho(\ell)^{-2} \ell^{-4}\propto \ell^{2p-4}\propto \ell^0$. Then its average value is $\langle A \rangle= A_0$. The turbulent therm only matters if $\beta=0$, hence the turbulent energy does not scale, that is why $\langle T \rangle= T_0$. 

To sum up, in the physical regime far from the core (i.e. when its presence can be neglected) there exists a single solution for the density profile $\varrho(\ell)=\ell^{-2}$. The latter corresponds to three possible configurations for the energy balance as listed below.

\begin{itemize}
 \item Balance between the gravitational and thermal energy (Case 1):
 \begin{equation}
 \label{equation_energy_balance_far_from_the_core_A}
 4 - \langle G \rangle \approx 0~~,
 \end{equation}
which can be realized for trans- or subsonic flows, if the turbulent velocity scales with $\beta>0$.

 \item Balance between the gravitational, turbulent and thermal energies (Case 2): 
 \begin{equation}
 \label{equation_energy_balance_far_from_the_core_B}
 \langle T \rangle + 4 - \langle G \rangle \approx 0~~,
 \end{equation}
which can be realized for trans- or subsonic flows, if the turbulent velocity does not scale ($\beta=0$).

The energy of the fluid element $E_0$ remains undetermined both in Case 1 and Case 2. 

\item Balance between the gravitational and turbulent energy (Case 3):
\begin{equation}
	\label{equation_energy_balance_far_from_the_core_C}
\langle T \rangle - \langle G \rangle \approx 0~~, 
\end{equation}
which can be realized only for supersonic flows, if the turbulent velocity is independent of spatial scale ($\beta=0$). 

The energy of the fluid element in Case 3 can be either positive ($E_0 = \langle A \rangle > 0$; implying an unstable energy state), if the accretion flow is also supersonic, or approximately zero (implying a marginally bound energy state), if the accretion flow is trans- or subsonic.
\end{itemize}

\subsubsection{The regime near to the core}
\label{subsubsec-EVT_solution_near}

A solution in this regime was obtained in Paper II only for supersonic flows -- it is a free-fall case characterized by density profile $\varrho(\ell)=\ell^{-3/2}$ and an energy balance $A_0 - G_1\approx 0$. It was shown that the energy of the fluid element is zero and thus the state of the gas can be characterized as marginally bound.

This solution cannot be reproduced from virial analysis within our model. One can not neglect all terms containing $\ell_0$ in equations (\ref{equation_<T>}) and (\ref{equation_<W>2}) to obtain their simplified forms (\ref{equation_<T>_proxy}) and (\ref{equation_<W>2_proxy}), accordingly. These approximations are not valid in the whole range of scales near the core: $\ell\gtrsim\ell_0$. In regard to the upper limit of this range, one can speculatively look for a transitional regime described by the solution $p=3/2$ according to the last row in Table \ref{tab-1}. However, the assumption $\langle \ddot{I} \rangle \simeq 0$ breaks in the regime near to the core and hence the equation (\ref{equation_EVT-simpler-far-expl}) is not anymore valid.

\section{Discussion}
\label{Sec-discussion}

\subsection{Basis of the model and main results}
\label{Subsec-basis and main results}

In a series of three papers (Paper I \& II and this work) we aim at modelling of the \rhopdf{} of a self-gravitating isothermal turbulent fluid. This corresponds to late stages of the evolution of MCs when the dense part of the PDF transforms to PLT(s) \citep{KNW_11,Girichidis_ea_14,Jaupart_Chabrier_2020}. Our approach is statistical and analytical. We introduce an {\it ensemble of MCs} as an useful approach to describe a set of MCs with the same \rhopdf, size, boundary densities and temperature \citep{DVK_17}. Furthermore, we introduce the concept of {\it average representative object}. This is the member of the ensemble with the simplest geometry and physics: spherically symmetric, radially isotropic and in dynamical equilibrium (steady state) in regard to the motions of fluid elements and gas molecules. An equation for the density profile is derived from the equations of the medium written for the average representative object. Approximate solutions of this equation yield the density profile in two regimes (Paper I \& II): far from the cloud core ($\varrho(\ell)=\ell^{-2}$) and near to the cloud core ($\varrho(\ell)=\ell^{-3/2}$). In these two regimes we presumed supersonic turbulent {\it and} accretion flows. In the present study, we applied the EVT to derive the density profile from an extended model which allows for super-, trans- and subsonic flows. A solution turns out to be possible only in the regime far from the core, where the profile $\varrho(\ell)=\ell^{-2}$ is reproduced by use of this independent approach. 

The energy balance corresponding to this regime can be realized in three configurations (Case 1, 2 and 3). Case 1 and 2 are possible only for trans- or subsonic flows. In the former, gravity is balanced by the thermal energy, since the turbulent velocity depends on spacial scale ($\beta>0$) and, hence, is subdominant in the balance equation (\ref{equation_EVT-simpler-far-expl}). Thus one obtains: $\langle G \rangle \approx 4$. In Case 2 turbulence is important because it does not scale. One obtains the equation: $\langle T \rangle + 4 - \langle G \rangle \approx 0$, i.e. gravity is balanced by the turbulent and thermal energies. Unfortunately, both in Case 1 and Case 2, the energy of the fluid element $E_0$ cannot be determined since the energy balance obtained in Paper II applies only for supersonic flows.

Case 3 is possible for supersonic turbulent flows. The energy balance equation obtained in this work reads: $\langle T \rangle - \langle G \rangle \approx 0$, i.e. gravity is balanced by turbulence. Here one can use the energy balance equation from Paper II to compute the energy of the fluid element. There are two possibilities. If the accretion flow is also supersonic, then $E_0 \approx \langle A \rangle > 0$, i.e. the energy state is unstable. For trans- or subsonic accretion flows, $E_0 \approx 0$ which corresponds to  a marginally bound energy state. 

In the regime near to the core, the second derivative of the moment of inertia in the EVT cannot be neglected. The only safe conclusion here is that the state of the gas is marginally bound and the core vicinity is not virialized.

An interesting question arises here. Can we extend the scope of validity of the solutions obtained in Paper I and II also to trans- and subsonic turbulent and accretion flows?
	
As demonstrated in Paper II, Sec. 3.2, if the fluid element is near to the core, the leading term in equation (14) there must be of order $\ell^{-1}$ and hence the thermal term $2(-p)\ln(\ell)$ is not important. Then, only accretion can balance the gravity of the core. Therefore the solution $p=-3/2$ in the regime near to the core is valid not only for supersonic, but also for trans- or subsonic flows. 

The solution in the regime far from the core is of more interest in the context of the present work. Let us consider equation (11) in Paper II, Sec. 3.1. The thermal term is again $2(-p)\ln(\ell)$ and is neglected due to the supersonic coefficients $A_0$ and $T_0$. However, in case of trans- or subsonic flows, $A_0$ and $T_0$ are of order unity or less, and then the term containing $\ln(\ell)$ can be neglected only if the fluid element is very close to the outer cloud boundary, i.e. $\ell = 1-\epsilon$, where $\epsilon\ll1$. This is a very narrow range indeed, in contrast to the expectation that a ``far from the core'' regime should span at least one order of magnitude or more. A possible solution of this problem can be to renormalize the model (and its equations), shifting the cloud boundary at $l_{\rm c} \equiv (1-\epsilon)l_{\rm c}$ and then repeating all considerations. The removed shell with thickness $\epsilon$ is added accordingly to the matter located outside the cloud, which is assumed to obey radial symmetry (see Sec. \ref{Sec-Model}). Thus this operation does not change anything in the model; it can be repeated $n\gg1$ times until the outer cloud boundary reaches values like $(1-n\epsilon)\sim\ell_0$, i.e. until the fluid element in consideration approaches the regime ``near to the core'' where the gravity of the core becomes important. Applying the above procedure, one can obtain the solution for the density profile $\varrho(\ell)=\ell^{-2}$ for trans- or subsonic flows and the energy balance equations (12) and (13) in Paper II will be also valid.

By the same reasoning one can calculate the energy of the fluid element $E_0$ for trans- and subsonic flows in the context of this work. If the turbulent velocity scales ($\beta>0$) and thus turbulence is not important, then $E_0 \approx \langle A \rangle - 4$, because from Paper II one has $\langle A \rangle - \langle G \rangle \approx E_0$. If $\beta=0$ and the turbulence matters, then again $E_0 \approx \langle A \rangle - 4$, because from Paper II one has $\langle A \rangle + \langle T \rangle - \langle G \rangle \approx E_0$, accordingly. In both cases the dimensionless accretion energy $\langle A \rangle \lesssim 1$ and one is to expect that $E_0$ is negative or at least approximately zero. Therefore the outer shells of the cloud should be stable or marginally bound.

\subsection{The model framework in the context of previous studies}
\label{Subsec-link to previous studies}

The presented results are linked to those from several previous studies. The case of collapsing homogeneous gas ball without accretion was originally investigated by \cite{Larson_69} and \cite{Penston_69a}, with self-gravity and isothermal gas pressure being the main acting forces. The numerical solutions of the equations of the medium yield a density profile with a slope of $p=2$ in the outer shells. \cite{Shu_77} and \cite{Hunter_77} also studied the problem analytically and obtained two density profiles: $p=2$ for the outer shells (but in static equilibrium: pressure provides support against gravity) and $p=3/2$ for the free-falling inner shells near to the singularity. The latter solution describes the so called ``inside-out collapse''. Later, \citet{Whitworth_Summers_85} have shown that the solutions of Larson-Penston and of Shu-Hunter are just the opposing ends of a 2-parameter family of solutions: for intrinsically unstable and initially quasi-hydrostatic clouds, respectively. This general picture was reproduced by the numerical study of \citet{Ogino_ea_99} who constrained its applicability to an epoch of accretion rate which is constant in time. (For stability analysis of the solutions, see also \citet{Hanawa_Nakayama_97}.)

Using numerical simulations, \cite{N-R_ea_15} explore a collapsing core embedded in a larger medium (called `cloud') and accreting material from it. They also obtain a density profile $p=2$ in the outer shells of the core as the latter collapse with the cloud. \cite{Li_18} derives a density profile with slope $p=2$ taking into account the interaction of gravity, accretion and turbulence in the course of the collapse. He asserts that the isothermal state is not a necessary condition for such profile -- the slope is universal in case of a scale-free gravitational collapse. Recently \cite{Jaupart_Chabrier_2020} develop the approach of \cite{Pan_ea_18,Pan_ea_19} and derive an analytical theory describing the PDF of density fluctuations in supersonic turbulent star-forming clouds in the presence of gravity. The \rhopdf{} is considered as a dynamical -- rather than stationary -- entity. The authors demonstrate the appearance of two PLTs in the \rhopdf{}, with two characteristic exponents, corresponding to different evolutionary stages in regard to the balance between turbulence and gravity. It is argued that a second PLT, with a characteristic slope of $-1.5$ ($p=2$), is a signature of regions in free-fall collapse.

A number of observations also confirm our result for the regime far from the core. The radial density profile in star cluster-forming molecular clumps is found to be very close to $p=2$ \citep{Mueller_ea_02,Evans_03,Wyr_ea_12,Palau_ea_14,Wyr_ea_16,Csengeri_ea_17,Zhang-Li_17}. Similar results, with some larger spread of $p$, have been obtained for the envelopes of prestellar/protostellar cores whereas the central region (of small relative volume) exhibits a flat profile \citep[e.g.,][]{Chandler_Richer_00, Shirley_ea_00, Motte_Andre_01, Smith_ea_11}, like in our model. This is generally consistent with the shape of stable or unstable Bonnor-Ebert spheres.

\subsection{Plausibility of the results}
\label{Subsec-Plausibility}

At least two issues are to be discussed in regard to the plausibility of the obtained solutions.

First, Cases 2 and 3 for the equations (\ref{equation_energy_balance_far_from_the_core_B}) and (\ref{equation_energy_balance_far_from_the_core_C}) of energy balance of the fluid element require lack of scaling ($\beta=0$) of the turbulent velocity fluctuations. An essential feature of both subsonic and supersonic turbulence is the positive power-law scaling relation with the size of considered volume. The lack of such scaling is to be interpreted rather like an additional sound speed but due to a macro-motion of fluid elements while thermal motion occurs at micro-scales. But although $\beta=0$ leads to a caveat, several observational works testify for such phenomenon. \cite{Goodman_1998} found lack of scaling of the nonthermal velocity dispersion in the so called ``coherent dense cores'' (extracted by isodensity contours of the ${\rm NH}_3$ emission): low-mass ($1-10~{\rm M}_{\odot}$) star-forming clumps with typical sizes of $\sim 0.1~{\rm pc}$. Similarly, \cite{Caselli_ea_02} obtained a constant nonthermal velocity dispersion in the interiors of dense low-mass cores extracted from ${\rm N}_2 {\rm H}^{+}$ maps. \cite{Pineda_ea_10} studied a sample of dense ${\rm NH}_3$ cores, in whose inner parts the nonthermal velocity dispersion does not depend on scale. \cite{Gibson_ea_09} observed dense high-mass CS cores wherein no velocity scaling is present. Analysing a sample of dense high-mass ($\sim 10^2 - 10^3~{\rm M}_{\odot}$) clumps delineated through HCN and CS isodensity contours, with sizes of several tenths of pc, \cite{Wu_ea_10} found a very weak scaling of velocity. By use of a different clump-finding technique ({\sc Gaussclumps}), \cite{Veltchev_ea_18} extracted high-mass ($\sim 10 - 10^3~{\rm M}_{\odot}$) CO condensations of low density and sizes between $0.1$ and $1$ pc in the star-forming region Rosette, which exhibit no ($\beta=0$) or a very weak scaling ($\beta=0.2$; see their Fig. 7).

A second issue is the condition for trans- or subsonic flows in Cases 1 and 2 (equations for energy balance \ref{equation_energy_balance_far_from_the_core_A} and \ref{equation_energy_balance_far_from_the_core_B}, respectively). The Larson-Penston solution for isothermal self-gravitating sphere \citep{Larson_69, Penston_69a} predicts that in case of established density profile $\varrho=\ell^{-2}$ at later stages of the collapse the velocity (accretion) field tends to a supersonic value ($\sim 3.3 c_{\rm s}$). \cite{GG_VS_20} find that the turbulent velocity appears to approach a constant fraction of the infall speed and thus, if the infall speed is sufficiently supersonic, so should be the turbulent velocity. Then the found solutions in Cases 1 and 2 do not matter. Indeed, the dense high-mass cores observed by \citet{Gibson_ea_09} and \citet{Wu_ea_10} exhibit supersonic nonthermal velocity dispersions. However, dense ``coherent cores'' \citep{Goodman_1998, Caselli_ea_02, Pineda_ea_10} and the high-mass cores of low density extracted by \cite{Veltchev_ea_18} are characterized by trans- or subsonic nonthermal velocity dispersions.

To sum up, the lack of scaling of turbulent velocity in Cases 2 and 3 and the condition for trans- or subsonic flows in Cases 1 and 2 may cause concerns how plausible are the obtained results. Nevertheless, there are several observational studies indicating that the solutions from our model are valid for several different types of cores. Case 2 may apply to dense ``coherent cores'' \citep{Goodman_1998,Caselli_ea_02,Pineda_ea_10} and to massive CO clumps associated with the low-density envelopes of star-forming cores \citep{Veltchev_ea_18}. On the other hand, Case 3 is similar to the dense high-mass objects studied by \citet{Gibson_ea_09} and \citet{Wu_ea_10}.

It should be acknowledged, however, that two of those datasets (those from \citealt{Gibson_ea_09} and \citealt{Wu_ea_10}), along with several others, were discussed by \citet{BP_ea_2011a} as possible evidence that the line-widths are actually reflecting infall motions, rather than turbulence, so that the velocity dispersion obeys a scaling: $\sigma_{\rm v} \sim \sqrt{\Sigma \ell}$, where $\Sigma$ is the column density, rather than a simple turbulent scaling of the form: $\sigma_{\rm v} \sim \ell^{\beta}$ (see Figs. 1 and 2 of \citealt{BP_ea_2011a}).

\subsection{Caveats}
\label{Subsec-caveats}

We recognize that our model is idealized and emphasize this mostly in view of four possible caveats. 

First, in regard to the cloud geometry, spherical symmetry is not the exact shape of clumps and cores in MCs. However, a 3D object that is characterized by a single spatial size can be reconstructed from the volume weighted \rhopdf, if and only if, the latter is a monotonic function. In such case a given scale as defined in equation (\ref{equation_def_scale}) can be linked to the density and thus the necessary one to one relation between density and radial position is generated.  

The choice of a sphere is appropriate since spherical geometry is a first-order approximation for a system without magnetic field and rotation. Moreover, the pressure forces tend to isotropize the system and its lowest-energy state corresponds to spherical shape. Also, a small deviation from the spherical symmetry does not cause significant changes in the results from the model. The considered average representative object is averaged in time and thus turbulence, which is homogeneous and isotropic by assumption, is not inconsistent with the spherical symmetry. We remind the reader that our aim is not to reproduce the complete dynamics and morphology of the fluid in an individual cloud, but rather to assess the general characteristics of a whole MC ensemble. 

A second caveat -- directly linked to the assumption of spherical symmetry -- might be the neglection of rotation and magnetic fields in our model. In the first case, angular momentum conservation should be taken into account and the state of minimum energy is a rotationally supported disk and not a sphere. However, the typical sizes of disks around protostars (tens to a few hundreds of AU) are much smaller than the spatial scales at which our model operates. Such disks form inside the cloud core, while outside it the velocity component (and the corresponding energy term) due to radial inflow dominates by far. Inclusion of magnetic fields in the model would also break the spherical symmetry and change the energy balance equation. Such construction would be justified in regard to an earlier epoch of cloud evolution characterized by possible magnetic support against collapse \citep{Shu_Adams_Lizano_87} while the pronounced PLT of the \rhopdf{} in our study implies advanced stages of self-gravitating clouds. To sum up, although we recognize the importance of rotation and magnetic fields for a realistic description of cloud dynamics, their neglection cannot affect significantly the density profile at the considered spatial scales and evolutionary time of the cloud.

The third possible caveat concerns the assumption of steady state. Indeed, a steady state or an approach to it has been suggested by various authors \citep[e.g.][]{Murray_Chang_15, Murray_ea_17, GG_VS_20, Gomez_ea_2021}. We note also that a quasi-steady state is not inconsistent with the collapse, as long as there is an external mass reservoir that supplies the accretion (implicitly assumed in our model). On the other hand, many authors consider MCs as dynamically evolving objects which are never in a steady state \citep[see][and the references there in]{VS_ea_19}. We assert that our assumption is justified as long as the final stages of MC evolution are considered, when the high-density part of the PDF consists of (a) well developed PLT(s). Numerical simulations show that the structures corresponding to this/these PLT(s) are in steady state \citep{KNW_11, Girichidis_ea_14, Khullar_ea_21}. 

The fourth caveat arises from the requirement for lack of turbulent-velocity scaling ($\beta=0$) in Cases 2 and 3. It seems strange in view of the typical picture in (subsonic and supersonic) turbulent media. Nevertheless, as discussed in the Sec. \ref{Subsec-Plausibility}, there exist several observational examples for such phenomenon which suggest that it is still not completely understood. \cite{Goodman_1998} and \cite{Pineda_ea_10} found a jump of the nonthermal velocity dispersion at the boundary of their ``coherent cores'': it is significantly larger outside them and scales positively with size. \citet{Goodman_1998} explain this behaviour with ``the marked decline in the coupling of the magnetic field to gas motions due to a decreased ion/neutral ratio in dense, high filling factor gas'' and hence the appearance of a dissipation threshold of MHD turbulence at the cores' edges. \citet{Pineda_ea_10} discuss the absence of velocity scaling and attribute it to possible changes in density, pressure and/or magnetic field at the cores' boundaries.

Last, we recognize that the presented model (Paper I \& II, and this work) is applicable only to the {\it protostellar} stage of MC evolution. The obtained solutions for the density ($\sim \ell^{-3/2}$) and velocity ($\sim \ell^{-1/2}$) profiles in the regime near to the core are monotonically increasing functions. This behaviour of the flow characterizes the protostellar cores while in the pre-stellar case the density is constant while the velocity decreases, in the region near to the centre, where the singularity (protostar) does not form yet \citep{Whitworth_Summers_85}. It should be noted though, that the {\it approach} used to derive the solution in the regime far from the core can be applied to the outer regions of pre-stellar cores. In Paper I \& II, the conditions to obtain this solution were: i) a steady state, and ii) a negligible effect of the gravity of the central region on a moving fluid element in the outer regions. In the framework in this Paper, one should add a requirement for negligible internal energy of  the central regions and for negligible change of the moment of inertia (see Section \ref{subsec-EVT_model}, \ref{subsubsec-<T>_explicit}, and \ref{subsubsec-<W>_explicit}). All these conditions are satisfied in the case of pre-stellar core.

\subsection{On the correspondence between the profile $\varrho(\ell)=\ell^{-2}$ and the assumption of steady state}
\label{Subsec-one to one correspondence}

Finally, we note an important feature of the density profile obtained for the regime far from the core. Considering a spherically symmetric cloud which accretes matter radially from its surroundings, of a density profile of type $\varrho(\ell)=\ell^{-2}$ is equivalent to a steady state flow through the cloud shells. In Papers I \& II and in this work we derived a profile $\varrho(\ell)=\ell^{-2}$ in the regime far from the core if the steady state flow is assumed. 

Let us demonstrate that the opposite is true as well: an accreting cloud characterized by a density profile $\varrho=\ell^{-2}$ and approximate spherical symmetry should be in a steady-state. Let $v_{\rm nt}$ be the non-thermal velocity dispersion in the cloud dominated by the infall motions. Then, as shown by \citet{Heyer_ea_2009} and \citet{BP_ea_2011a},  $v_{\rm nt}^2\propto G \Sigma R$, where $\Sigma$ is the surface mass density and $R$ is the radius of the considered shell (i.e., $R$ corresponds to $\equiv\ell$). Combining this relation with the one for the density profile and taking into account that $\Sigma \propto \varrho\ell \propto \ell^{-1}$, one obtains $v_{\rm nt}= {\rm const}(\ell)$. On the other hand, the continuity equation in spherical coordinates reads:

\[ \frac{\partial \varrho}{\partial t} + \vec{\triangledown} \cdot (\varrho \vec{v}) =\frac{\partial \varrho}{\partial t} + \frac{1}{\ell^2} \frac{\partial}{\partial\ell} (\ell^2 \varrho v_{\rm nt}) = 0~,\]
where we assume for the averaged velocity: $\vec{v} \approx (v_{\rm nt},0,0)$. Using the considerations above, it is obvious that $\ell^2 \varrho v_{\rm nt} = {\rm const}(\ell)$. Hence $\partial\varrho/\partial t = 0$, i.e. the density field does not depend explicitly on the time. Thus the cloud is in approximate steady state.

The equivalence between the assumption of steady state and the density profile $\varrho=\ell^{-2}$ has been also shown by \citet{Li_18}. Recently \citet{Gomez_ea_2021} demonstrated as well that such slope is an attractor -- other slopes approach this value in the course of the cloud evolution. This is also consistent with the obtained possibility in the case of trans- or subsonic flow for a negative total energy of the fluid element in the regime far from the core in our model which implements stability of the outer shells.

\section{Conclusions}
\label{Sec-conclusions}

In this study we extend the model of ensembles of self-gravitating turbulent clouds developed in Papers I and II, applying the Eulerian Virial Theorem (EVT) and broadening the scope of consideration to include super-, trans- and subsonic flows. Our analysis shows that the solution $\varrho(\ell)=\ell^{-2}$ is confirmed in the regime far from the core. Moreover, the obtained energy balance equations yield the value of the energy of the fluid element which remains indeterminate in our original approach, by use of the differential equations of the medium (Papers I and II). This energy can be negative in the cases of trans- or subsonic flows, i.e. the energy balance in the solution far from the core shows that the outer cloud shells can be stable. Thus this solution for the density profile might correspond to a stable dynamical state and must be observable. In the regime near to the core, one cannot derive an equation for the density profile, this is because the second derivative of the moment of inertia of the gas could not be neglected. In this regime we arrive only at the qualitative conclusion that the gas near to the core is not virialized. Also, the energy of the fluid element in this regime is close to zero which suggests that the gas state can be characterized as marginally bound. In this way, the model is further developed by use of the EVT as an independent approach.

\section*{Acknowledgements}
The authors are thankful to the anonymous referee whose remarks and suggestions helped us to substantially improve this paper. 
S.D. and T.V. acknowledge support by the DFG (Deutsche Forschungsgemeinschaft) under grant KL 1358/20-3. T.V. thanks for the additional funding from the Ministry of Education and Science of the Republic of Bulgaria, National RI Roadmap Project DO1-176/29.07.2022. R.S.K. acknowledges support from the Deutsche Forschungsgemeinschaft (DFG) via the collaborative research centre (SFB 881, Project-ID 138713538) ``The Milky Way System'' (sub-projects A1, B1, B2 and B8) and from the Heidelberg cluster of excellence (EXC 2181 - project-ID 390900948) ``STRUCTURES: A unifying approach to emergent phenomena in the physical world, mathematics, and complex data''. He also thanks for funding from the European Research Council in the ERC synergy grant ``ECOGAL – Understanding our Galactic ecosystem: From the disk of the Milky Way to the formation sites of stars and planets'' (project-ID 855130).

\section*{Data availability}
No new data were generated or analysed in support of this research.

\bibliographystyle{mnras}

\begin{thebibliography}{}
\makeatletter
\relax
\def\mn@urlcharsother{\let\do\@makeother \do\$\do\&\do\#\do\^\do\_\do\%\do\~}
\def\mn@doi{\begingroup\mn@urlcharsother \@ifnextchar [ {\mn@doi@}
  {\mn@doi@[]}}
\def\mn@doi@[#1]#2{\def\@tempa{#1}\ifx\@tempa\@empty \href
  {http://dx.doi.org/#2} {doi:#2}\else \href {http://dx.doi.org/#2} {#1}\fi
  \endgroup}
\def\mn@eprint#1#2{\mn@eprint@#1:#2::\@nil}
\def\mn@eprint@arXiv#1{\href {http://arxiv.org/abs/#1} {{\tt arXiv:#1}}}
\def\mn@eprint@dblp#1{\href {http://dblp.uni-trier.de/rec/bibtex/#1.xml}
  {dblp:#1}}
\def\mn@eprint@#1:#2:#3:#4\@nil{\def\@tempa {#1}\def\@tempb {#2}\def\@tempc
  {#3}\ifx \@tempc \@empty \let \@tempc \@tempb \let \@tempb \@tempa \fi \ifx
  \@tempb \@empty \def\@tempb {arXiv}\fi \@ifundefined
  {mn@eprint@\@tempb}{\@tempb:\@tempc}{\expandafter \expandafter \csname
  mn@eprint@\@tempb\endcsname \expandafter{\@tempc}}}

\bibitem[\protect\citeauthoryear{Ballesteros-Paredes et al.}{2011}]{BP_ea_2011a}
Ballesteros-Paredes, J., Hartmann, L. W., V\'azquez-Semadeni, E., Heitsch, F., Zamora-Avil\'es, M. A., 2011, MNRAS, 411, 65
\bibitem[\protect\citeauthoryear{Ballesteros-Paredes}{2006}]{BP_06}
Ballesteros-Paredes, J., 2006, MNRAS, 372, 443
\bibitem[\protect\citeauthoryear{Caselli et al.}{2002}]{Caselli_ea_02}
Caselli, P., Benson, P. J., Myers, P. C., Tafalla, M. 2002, ApJ, 572, 238
\bibitem[\protect\citeauthoryear{Chandler \& Richer}{2000}]{Chandler_Richer_00}
Chandler, C., Richer, J., 2000, ApJ, 530, 851
\bibitem[\protect\citeauthoryear{Chen et al.}{2020}]{Chen_ea_20}
Chen H. H.-H., Offner S. S. R., Pineda J. E., Goodman A. A., Burkert A., Ginsburg A., Choudhury S., 2020, arXiv e-prints, p. arXiv:2006.07325
\bibitem[\protect\citeauthoryear{Csengeri et al.}{2017}]{Csengeri_ea_17}
Csengeri T., et al., 2017, A\&A, 600, L10
\bibitem[\protect\citeauthoryear{Donkov, Veltchev \& Klessen}{2017}]{DVK_17}	
Donkov, S., Veltchev, T., Klessen, R. S., 2017, MNRAS, 466, 914
\bibitem[\protect\citeauthoryear{Donkov \& Stefanov}{2018}]{DS_18}	
Donkov, S., Stefanov, I. Zh., 2018, MNRAS, 474, 5588 (Paper I)
\bibitem[\protect\citeauthoryear{Donkov \& Stefanov}{2019}]{DS_19}	
Donkov, S., Stefanov, I. Zh., 2019, MNRAS, 485, 3224 (Paper II)
\bibitem[\protect\citeauthoryear{Elmegreen \& Scalo}{2004}]{Elme_Scalo_04}
Elmegreen, B., Scalo, J., 2004, ARA\&A, 42, 211E
\bibitem[\protect\citeauthoryear{Evans}{2003}]{Evans_03}
Evans II N., 2003, in Curry C. L., Fich M., eds, SFChem 2002: Chemistry as a Diagnostic of Star Formation. p. 157
\bibitem[\protect\citeauthoryear{Federrath et al.}{2010}]{Federrath_ea_10}
Federrath, C., Roman-Duval, J., Klessen, R., Schmidt, W., Mac Low, M.-M., 2010, A\&A, 512, 81
\bibitem[\protect\citeauthoryear{Ferriere}{2001}]{Ferriere_01}
Ferriere, K. M. 2001, Reviews of Modern Physics, 73, 1031
\bibitem[\protect\citeauthoryear{Ganguly et al.}{2022}]{Ganguly_ea_22}
Ganguly, S., Walch, S., Clarke S.D., Seifried, D., arXiv: 2204.02511
\bibitem[\protect\citeauthoryear{Gibson et al.}{2009}]{Gibson_ea_09}
Gibson, D., Plume, R., Bergin, E., Ragan, S., Evans, N. 2009, ApJ, 705, 123
\bibitem[\protect\citeauthoryear{Girichidis et al.}{2014}]{Girichidis_ea_14}
Girichidis, P., Konstandin, L., Whitworth, A., Klessen, R., 2014, ApJ, 781, 91
\bibitem[\protect\citeauthoryear{Goodman et al.}{1998}]{Goodman_1998}
Goodman, A. A., Barranco, J. A., Wilner, D. J., Heyer, M. H., 1998, ApJ, 504, 223-246
\bibitem[\protect\citeauthoryear{Gomez et al.}{2021}]{Gomez_ea_2021}
Gomez, G., V\'azquez-Semadeni, E., Palau, A., 2021, MNRAS, 502, 4963
\bibitem[\protect\citeauthoryear{Guerrero-Gamboa \& V\'azquez-Semadeni}{2020}]{GG_VS_20}
Guerrero-Gamboa, R., V\'azquez-Semadeni, E., 2020, ApJ, 903, 136
\bibitem[\protect\citeauthoryear{Hanawa \& Nakayama}{1997}]{Hanawa_Nakayama_97}
Hanawa, T., Nakayama, K., 1997, ApJ, 484, 238 
\bibitem[\protect\citeauthoryear{Hennebelle \& Falgarone}{2012}]{HF_12}
Hennebelle, P. \& Falgarone, E., 2012, A\&ARv, 20, 55H
\bibitem[\protect\citeauthoryear{Heyer et al.}{2009}]{Heyer_ea_2009}
Heyer, M., Krawczyk, C., Duval, J., Jackson, J. M., 2009, ApJ, 699, 1092
\bibitem[\protect\citeauthoryear{Hunter}{1977}]{Hunter_77}
Hunter, C. 1977, ApJ, 218, 834
\bibitem[\protect\citeauthoryear {Jaupart \& Chabrier}{2020}]{Jaupart_Chabrier_2020}
Jaupart, E., Chabrier, G., 2020, ApJL, 903, L2
\bibitem[\protect\citeauthoryear{Keto}{2020}]{Keto_20}
Keto, E., MNRAS, 493, 2834
\bibitem[\protect\citeauthoryear{Khullar et al.}{2021}]{Khullar_ea_21}
Khullar, S., Federrath, C., Krumholz, M. R., Matzner, C. D., 2021, MNRAS, accepted, arxiv:2107.00725
\bibitem[\protect\citeauthoryear{Klessen \& Glover}{2016}]{KG_16}
Klessen, R. S., \& Glover, S. C. O., 2016, Star Formation in Galaxy Evolution:
Connecting Numerical Models to Reality, Saas-Fee Advanced Course
\bibitem[\protect\citeauthoryear{Kolmogorov}{1941}]{K41}
Kolmogorov, A., 1941, Dokl. Akad. Nauk SSSR, 30, 301
\bibitem[\protect\citeauthoryear{Kritsuk et al.}{2007}]{Kritsuk_ea_07}
Kritsuk, A., Norman, M., Padoan, P., \& Wagner, R., 2007, ApJ, 665, 416
\bibitem[\protect\citeauthoryear{Kritsuk, Norman \& Wagner}{2011}]{KNW_11}
Kritsuk, A., Norman, M., \& Wagner, R., 2011, ApJ, 727, L20
\bibitem[\protect\citeauthoryear{Krumholz}{2015}]{Krh_15}
Krumholz, M., R., 2015, {\it Notes on Star Formation} (arXiv: 1511.03457)
\bibitem[\protect\citeauthoryear{Larson}{1969}]{Larson_69}
Larson, R., 1969, MNRAS, 145, 271
\bibitem[\protect\citeauthoryear{Larson}{1981}]{Larson_81}
Larson, R., 1981, MNRAS, 194, 809
\bibitem[\protect\citeauthoryear{Li}{2018}]{Li_18}
Li, G.-X., 2018, MNRAS, 477, 4951L
\bibitem[\protect\citeauthoryear{McKee \& Zweibel}{1992}]{MZ_92}	
McKee, Christopher F., Zweibel, Elen G., 1992, ApJ, 399, 551M
\bibitem[\protect\citeauthoryear{Molina et al.}{2012}]{Molina_ea_12}
Molina, F., Glover, S. C. O., Federrath, C., Klessen, R. S., 2012, MNRAS, 423, 2680
\bibitem[\protect\citeauthoryear{Motte \& Andr\'e}{2001}]{Motte_Andre_01}
Motte, F., Andr\'e, P., 2001, A\&A, 365, 440
\bibitem[\protect\citeauthoryear{Mueller et al.}{2002}]{Mueller_ea_02}
Mueller K. E., Shirley Y. L., Evans II N. J., Jacobson H. R., 2002, ApJS, 143, 469
\bibitem[\protect\citeauthoryear{Murray \& Chang}{2015}]{Murray_Chang_15}
Murray, N., Chang, Ph., 2015, ApJ, 804, 44 
\bibitem[\protect\citeauthoryear{Murray et al.}{2017}]{Murray_ea_17}
Murray, D., Chang, Ph., Murray, N., Pittman, J., 2017, MNRAS, 465, 1316 
\bibitem[\protect\citeauthoryear{Naranjo-Romero, V\'azquez-Semadeni  \& Loughnane}{2015}]{N-R_ea_15}
Naranjo-Romero R., V\'azquez-Semadeni, E., Loughnane R. M., 2015, ApJ, 814, 48
\bibitem[\protect\citeauthoryear{Ogino, Tomisaka \& Nakamura}{1999}]{Ogino_ea_99}
Ogino, S., Tomisaka, K., Nakamura, F., 1999, PASJ, 51, 637
\bibitem[\protect\citeauthoryear{Padoan et al.}{2006}]{Pad_ea_06}
Padoan, P., Juvela, M., Kritsuk, A., Norman, M., 2006, ApJ, 653, L125
\bibitem[\protect\citeauthoryear{Palau et al.}{2014}]{Palau_ea_14}
Palau A., et al., 2014, ApJ, 785, 42
\bibitem[\protect\citeauthoryear{Pan et al.}{2018}]{Pan_ea_18}
Pan L., Padoan P., Nordlund A., 2018, ApJ, 866, L17
\bibitem[\protect\citeauthoryear{Pan et al.}{2019}]{Pan_ea_19}
Pan L., Padoan P., Nordlund A., 2019, ApJ, 881, 155
\bibitem[\protect\citeauthoryear{Penston}{1969}]{Penston_69a}
Penston, M., V., 1969a, MNRAS, 145, 457
\bibitem[\protect\citeauthoryear{Pineda et al.}{2010}]{Pineda_ea_10}
Pineda J. E., Goodman A. A., Arce H. G., Caselli P., Foster J. B., Myers P. C., Rosolowsky E. W., 2010, ApJ, 712, L116
\bibitem[\protect\citeauthoryear{Riley, Hobson \& Bence}{2006}]{RHB_2006}
Riley, K. F., Hobson, M. P., Bence, S. J., 2006, Mathematical Methods for Physics and Engineering. Cambrige University Press, Cambridge, New York
\bibitem[\protect\citeauthoryear{Shetty et al.}{2012}]{Shetty_2012}
Shetty, R., Beaumont, C. N., Burton, N. G., Kelly B. C., Klessen R. S., 2012, MNRAS, 425, 1, 720-729
\bibitem[\protect\citeauthoryear{Shirley et al.}{2000}]{Shirley_ea_00}
Shirley, Evans II, N. J., Rawlings, J., Gregersen, E., ApJS, 131, 249
\bibitem[\protect\citeauthoryear{Shu}{1977}]{Shu_77}
Shu, F. H. 1977, ApJ, 214, 488
\bibitem[\protect\citeauthoryear{Shu, Adams \& Lizano}{1987}]{Shu_Adams_Lizano_87}
Shu, F. H., Adams, F., Lizano, S., 1987, ARA\&A, 25, 23
\bibitem[\protect\citeauthoryear{Smith et al.}{2011}]{Smith_ea_11}
Smith, R. J., Glover, S. C. O., Bonnell, I. A., Clark, P. C., Klessen, R. S., 2011, MNRAS, 411, 1354
\bibitem[\protect\citeauthoryear{V\'azquez-Semadeni}{1994}]{VS_94}
V\'azquez-Semadeni, E., 1994, ApJ, 423, 681
\bibitem[\protect\citeauthoryear{V\'azquez-Semadeni}{2010}]{VS_10}
V\'azquez-Semadeni, E., 2010, in Kothes, R., Landecker, T., Willis, A., ASP Conf. Ser. Vol. 438, The Dynamic Interstellar Medium: A Celebration of the Canadian Galactic Plane Survey. Astron. Soc. Pac., San Francisco, p. 83; arXiv 1009.3962
\bibitem[\protect\citeauthoryear{V\'azquez-Semadeni et al.}{2019}]{VS_ea_19}
V\'azquez-Semadeni, E., Palau, A., Ballesteros-Paredes, J., G\'omez, G., Zamora-Aviles, M., 2019, MNRAS, 490, 3061
\bibitem[\protect\citeauthoryear{Veltchev et al.}{2018}]{Veltchev_ea_18}
Veltchev, T., Ossenkopf-Okada, V., Stanchev, O., Schneider, N., Donkov, S., Klessen, R. S. 2018, MNRAS, 475, 2215.
\bibitem[\protect\citeauthoryear{Veltchev et al.}{2019}]{Veltchev_ea_19}
Veltchev, T., Girichidis, P., Donkov, S., Schneider, S., Stanchev, O., Marinkova, L., Seifried, D., Klessen, R. S., 2019, MNRAS, 489, 788
\bibitem[\protect\citeauthoryear{Whitworth \& Summers}{1985}]{Whitworth_Summers_85}
Whitworth, A., Summers, D., 1985, MNRAS, 214, 1
\bibitem[\protect\citeauthoryear{Wu et al.}{2010}]{Wu_ea_10}
Wu, J., Evans, N. J., Shirley, Y. L., Knez, C. 2010, ApJS, 188, 313
\bibitem[\protect\citeauthoryear{Wyrowski et al.}{2012}]{Wyr_ea_12}
Wyrowski F., Gusten R., Menten K. M., Wiesemeyer H., Klein B., 2012, A\&A, 542, L15
\bibitem[\protect\citeauthoryear{Wyrowski et al.}{2016}]{Wyr_ea_16}
Wyrowski F., et al., 2016, A\&A, 585, A149
\bibitem[\protect\citeauthoryear{Zhang \& Li}{2017}]{Zhang-Li_17}
Zhang C.-P., Li G.-X., 2017, MNRAS, 469, 2286
\bibitem[\protect\citeauthoryear{Zhivkov}{1999}]{Zhivkov_99}
Zhivkov, A., 1999, Differential Equations Problems. Demokratichni tradicii - Demetra, Sofia

\makeatother
\end{thebibliography}

\end{document}